\documentclass[pre,aps,longbibliography,amsmath,amsfonts,amssymb,twocolumn]{revtex4-1}
\usepackage{graphicx,CJK}
\newcommand{\void}{\varnothing}
\newcommand{\betae}{1.020(5)}
\newcommand{\betaer}{1.020(5)}
\newcommand{\nupee}{2.04(1)}
\newcommand{\nupae}{3.55(2)}
\begin{document}
\begin{CJK*}{UTF8}{mj}
\title{Order-parameter critical exponent of absorbing phase transitions in one-dimensional systems with two symmetric absorbing states}
	\author{Su-Chan Park (박수찬)}
\affiliation{The Catholic University of Korea, Bucheon 14662, Republic of Korea}
\begin{abstract}
Via extensive Monte Carlo simulations along with systematic analyses of corrections to scaling, we estimate the order parameter critical exponent $\beta$ of absorbing phase transitions in systems with two symmetric absorbing states.
The value of $\beta$ was conjectured to be $\frac{13}{14}\approx 0.93$ and Monte Carlo simulation studies in the literature have repeatedly reproduced  values consistent with the conjecture. 
In this paper, we systematically estimate $\beta$ by analyzing the effective exponent after finding how strong corrections to scaling are.
We show that the widely accepted numerical value of $\beta$ is not correct. Rather, we obtain $\beta = \betaer$ from different models with two symmetric absorbing states.
\end{abstract}
\date{\today}
\maketitle
\end{CJK*}
\section{\label{Sec:intro}Introduction}
Absorbing phase transitions have been extensively studied and 
have played an important role in figuring out 
a theory as to what determines the universal critical behavior in nonequilibrium 
systems~\cite{H2000,O2004}. To find critical exponents accurately is probably the most 
important starting point of formulating such a theory, unless an analytic method is available.
In this regard, the controversy~\cite{SB2008,Park2014PRE} as to whether 
the pair contact process with diffusion~\cite{HH2004} belongs to the directed percolation universality class or not is a vivid example of the importance of accurate values.
Up to now, many universality classes have been found in absorbing phase transitions, but
 critical exponents with high accuracy are available only for a small fraction of the classes; a notable example is the directed percolation (DP) class in one dimension~\cite{J1999}. 

Based on numerical simulations, in the mean time, Jensen~\cite{J1994} conjectured 
exact rational numbers about the critical exponents 
of systems with two symmetric absorbing states, which would in turn suggest a solvability.
We would like to mention in passing that the universality class with two symmetric absorbing states 
is referred to in several ways, such as 
parity conserving class~\cite{GKvdT1984,Jensen1993JPA}, directed Ising (DI) class~\cite{HKPP1998},
$Z_2$-symmetric directed percolation or DP2~\cite{H2000} class, and generalized voter class~\cite{HCDM2005}.
In this paper, we will use the term DI class.

Recent extensive simulations~\cite{Park2013}, however, refuted  
the conjecture, though some exponents such as $\beta/\nu_\perp$ and $\eta$
still remain within the conjecture ($\frac{1}{2}$ and $0$, 
respectively). In particular, it remains an open question 
to understand why $\eta$ seems to be exactly zero~\cite{Grassberger2013,Araujo2014}.

Although most exponents of the DI class have been found with accuracy~\cite{Park2013,Grassberger2013},
an unequivocal estimate of the order parameter exponent $\beta$ still remains open.
In Ref.~\cite{J1994}, $\beta$ was conjecture to 
be $\frac{13}{14} \approx 0.93$, which has been repeatedly reproduced 
within error bars in the literature~\cite{J1994, KP1994, M1994, Zhong1995,HO1999,MO2000,Argolo2020}. 
However, many numerical analyses 
are based on a (naive) power-law fitting of the form $\Delta^\beta$, where 
$\Delta$ is the distance from the critical point. A power-law fitting is error-prone 
due to the presence of corrections to scaling, even if one happens to know
the exact critical point. 
More systematic studies analyzing the effective exponent are actually available~\cite{HO1999,MO2000}, 
but corrections to scaling are not analyzed systematically in these studies.

Since the conjecture by and large
turned out to be invalid, it is necessary to check
if $\beta$ remains within the conjecture just like $\eta$ and $\beta/\nu_\perp$. 
In this paper, we would like to answer this question by extensive Monte Carlo simulations.
To this end, we first introduce a systematic data-analysis method, including how to 
estimate corrections to scaling. 
Using the method, we find the order parameter exponent $\beta$.

The structure of this paper is as follows.
In Sec.~\ref{Sec:model}, we explain the models
and introduce quantities that we are interested in. In particular,
we suggest a method to analyze corrections to scaling, which allows
for a systematic estimate of $\beta$.
Section~\ref{Sec:result} presents simulation results. 
Then we summarize the work in Sec.~\ref{Sec:sum}
with a discussion about the implications of our conclusion.

\section{\label{Sec:model}Model and Methods}
In this paper, we study two kinds of one-dimensional models:
the branching annihilating random walks with $2m$ offspring with
positive integer $m$, which will be abbreviated as
BAW($2m$)~\cite{TT1992} and 
the contact process with two-particle branching and annihilation, which will be called CP2 in short~\cite{InuiT1998PRL}.
For the BAW($2m$), we consider two different cases: $m=1$ (two offspring) and $m=2$ (four offspring).
Note that BAW($2m$) and CP2 in one dimension can be mapped to models with two symmetric absorbing states and
are believed to belong to the DI class in one dimension.

For all the models, we consider a one-dimensional
lattice, whose size will be denoted by $L$, with periodic boundary conditions. 
Each site can accommodate at most one particle.
A site with (without) a particle will be denoted by $A$ ($\void$) and we will say that
this site is occupied (vacant).
The initial condition is always such that every site is occupied.

In the BAW($2m$), each particle can hop to one of its nearest neighbors with rate $p$.
The hopping is unbiased. That is, one of its nearest neighbors is chosen with equal probability in the hopping events.
With rate $1-p$, each particle can branch $2m$ offspring: $m$ offspring 
will be placed at $m$ consecutive sites on the right (left) hand side
of the branching particle.
If two particles are to occupy a same site by any attempt, the attempt 
is ignored with probability $1-q$ and with 
probability $q$ two particles at the same site are removed immediately ($A+A\rightarrow \void$), which 
makes all sites accommodate at most a single particle at all time. 

The rules of the CP2 can be represented as
\begin{align}
	\nonumber
	&AA \rightarrow \void\void\text{ with rate } 1,\\
	&\void \void A\text{ or } A\void\void \rightarrow AAA  \text{ with  rate } \lambda.
\end{align}
We note that the CP2 with the fully occupied initial condition is 
identical to the one-dimensional interacting monomers  
model with two symmetric absorbing states~\cite{PP2008PRE}, though they are different in
higher dimensions~\cite{Park2012PRE}.

In this paper, we are interested in the behavior of the density $\rho(t)$ of occupied sites at time $t$ and its steady state value $\rho_s$, defined as
\begin{align}
	\rho(t) \equiv \frac{1}{L} \sum_{i=1}^L \langle \sigma_i(t) \rangle,\quad
	\rho_s \equiv \lim_{t\rightarrow \infty} \rho(t),
\end{align}
where $\sigma_i(t) = 1 (0)$ if site $i$ is occupied (vacant) at time $t$, $\langle \ldots \rangle$ stands
for average over ensemble, and we implicitly assume the infinite $L$ limit for the definition of $\rho(t)$.

Close to the critical point, scaling behaviors of these quantities are expected to be
\begin{align}
	\rho(t) &= t^{-\delta} F\left (\Delta t^{1/\nu_\|} \right ),\nonumber \\
	\rho_s(\Delta) &= A \Delta^\beta \left [ 1 + B \Delta^\chi + o(\Delta^\chi)\right ],
	\label{Eq:scale_rhos}
\end{align}
where $\Delta$ is the distance from the critical point; $\beta$, $\nu_\|$, $\delta$ are 
critical exponents with the relation $\delta = \beta/\nu_\|$; 
$F$ is a universal scaling function; 
the exponent $\chi$ dictates the strength of the leading corrections to scaling;
$o(x)$ stands for all terms that decay faster than $x$  as $x\rightarrow 0$; and $A$, $B$ are constants.
We choose $\Delta$ to be positive when $\rho_s$ is nonzero, that is,
when the system is in an active phase.

The efficient and systematic way of finding critical point 
as well as the critical exponent $\delta$ is to analyze the effective exponent $-\delta_\text{e}$ defined as
\begin{align}
	-\delta_\text{e}(t;b) = \frac{\ln[\rho(t)] - \ln[\rho(t/b)]}{\ln b},
\end{align}
where $b$ is a constant larger than 1.
For ease of analysis, $i$-th measurement of the density is done at (integer time) $T_i$ defined as
\begin{align}
	T_i = \begin{cases} i, & \text{ for } 1\le i \le 40,\\
		\left \lfloor 40 \times 2^{(i-40)/15} \right \rfloor, & \text{ for } 
		41 \le i \le 55,\\
		2 T_{i-15}, & \text{ for } i \ge 56,
	\end{cases}
\end{align}
where $\lfloor x \rfloor$ is the floor function (greatest integer
that is not larger than $x$).
Note that the time span corresponding to 50 measurements for $t > 40$ is roughly one decade
on a logarithmic scale.
With this choice of measurement timing, the effective exponent 
can be easily calculated if we set $b = 2^{k}$ ($k=1,2,\ldots$).
To find the critical point as well as the critical exponent $\delta$, 
we need information about corrections to scaling, which 
can be obtained by a method suggested in Refs.~\cite{Park2013,Park2014PRE}.

After having the critical point, we estimate $\beta$ by analyzing the 
effective exponent  $\beta_\text{e}$, defined as
\begin{align}
	\beta_\text{e} ( \Delta; b) &\equiv \frac{\ln [ \rho_s(b \Delta)/\rho_s(\Delta) ]}{\ln b}\nonumber\\
	&= \beta + B \Delta^\chi \frac{b^\chi-1}{\ln b} + o(\Delta^\chi),
	\label{Eq:betae}
\end{align}
where $b$ is a constant.
The effective exponent can be studied systematically if
we get the stationary state density at regular values of $\Delta$
on a logarithmic scale. In practice, we simulated  systems at $\Delta = \Delta_n$, where
\begin{align}
	\Delta_n = 2^{n/2} \times 10^{-3}.
	\label{Eq:Deltan}
\end{align}
With this choice, we will set $b = 2^{k/2}$ ($k=1,2,\ldots$).

Unless $\Delta$ is sufficiently small, $\beta_\text{e}$ is significantly affected by corrections
to scaling that are characterized by the exponent $\chi$. Accordingly, an accurate estimate of $\beta$ from $\beta_\text{e}$ requires information 
about $\chi$. Using an idea similar to 
that in Ref.~\cite{Park2013,Park2014PRE},
we introduce the corrections-to-scaling function $Q$ defined as
\begin{align}
	Q(\Delta;b,\chi) = \frac{ \ln \rho_s(\Delta) + \ln \rho_s(b^2\Delta)-2 \ln \rho_s(b \Delta) }{(b^\chi-1)^2}.
	\label{Eq:Q}
\end{align}
If we have the correct value of $\chi$, $Q$ behaves in the 
asymptotic regime  ($\Delta \rightarrow 0$) as
$Q\approx B\Delta^\chi$ without dependence on $b$. This fact will
be used for a consistency check of the $\chi$ estimate. 

With the value of $\chi$, we analyze $\beta_\text{e}$ 
as a function of $\Delta^\chi$,
which should be a straight line for the small $\Delta$ regime.
Hence, we can find $\beta$ by a linear extrapolation.

\section{\label{Sec:result} Results}
In this section, we present simulation results.
Let us begin with the BAW(4) with $q=1$.
We first find the critical point by 
studying the effective exponent $-\delta_\text{e}$. 
We simulated a system of size $L=2^{23}$ up to time $t=T_{309}\approx 10^7$ and the density is averaged over $800$ (for $p=0.7219$) or $1000$
(for other cases) independent runs. 

\begin{figure}
	\includegraphics[width=\linewidth]{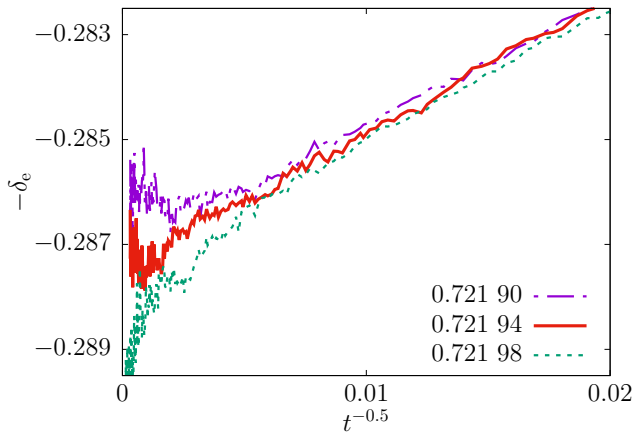}
	\caption{\label{Fig:4pc} Plots of $-\delta_\text{e}$ vs $t^{-0.5}$
	for $p=0.7219$, 0.721~94, and 0.721~98 (top to bottom) 
	for the BAW(4). Here, $b$ is set to 16.
	$\delta_\text{e}$ for $p=0.721~94$ approaches $0.2872$, 
	the value of the DI class~\cite{Park2013}, while the other 
	curves deviate from the straight line behavior as they approach the ordinate. Hence we conclude $p_c = 0.721~94(4)$, where the 
	number in parentheses indicates the uncertainty of the last 
	digit.
	}
\end{figure}

In Fig.~\ref{Fig:4pc}, we depict $-\delta_\text{e}$ with $b=16$ 
against $t^{-0.5}$ for $p=0.7219$, 0.721~94, and 0.721~98.
We would like to mention in passing that the leading behavior of 
corrections to scaling  was found to be $t^{-0.5}$ by the method
in Ref.~\cite{Park2014PRE} (details not shown here).
The curve for $p=0.7219$ (0.721~98) clearly veers up (down), 
indicating that the system is in the active (absorbing) phase.
Accordingly, we find $p_c = 0.721~94(4)$ and $\delta \approx 0.2872$. 
This critical point should be compared with that in Ref.~\cite{J1994},
which is $0.7215$.
Notice that the numerical value of $\delta$ estimated in Fig.~\ref{Fig:4pc} is consistent with the result in Ref.~\cite{Park2013}.

\begin{figure}[b]
	\includegraphics[width=\linewidth]{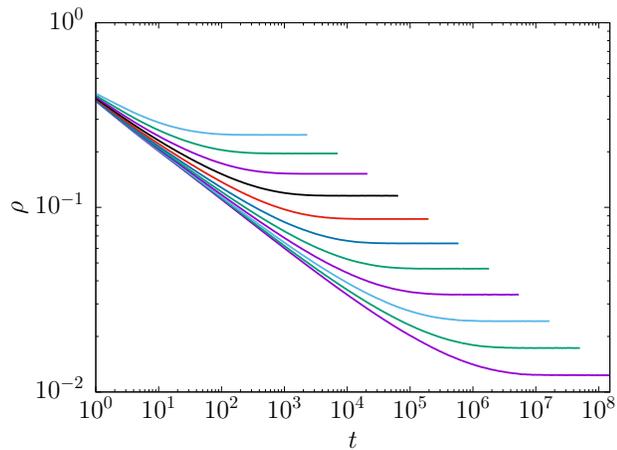}
	\caption{\label{Fig:SA4} Double logarithmic plots of $\rho$ vs $t$
	at $p = p_c -\Delta_n$ with $n=5,6,\ldots,15$ (bottom to top)
	for the BAW(4).
	$\Delta_n$ is defined in Eq.~\eqref{Eq:Deltan}.
	}
\end{figure}

Having the critical point $p_c$, we found the steady state density at 
$\Delta\equiv p_c-p=\Delta_n$.
We simulated systems of size $2^{20}$ or $2^{21}$,
which are large enough for the finite size effect not to affect the 
steady state value significantly. For better statistics,
we take the average over $1000$ ($n\ge 8$) or $2000$ ($n\le 7$) 
independent realizations.
In Fig.~\ref{Fig:SA4}, we present the resulting $\rho(t)$ for various $n$'s.

When we calculated the steady state density $\rho_s$, we took
an average over the last 25 measurement points, which correspond to a 
half-decade on a logarithmic scale. This averaging procedure to find the steady state
density is also applied to other models, to be studied soon.
We obtained $\rho_s$ with error  of
size $\sim 10^{-5}$ (the relative error for $n=5$ is about $10^{-3}$,
which is the largest in our data). 

\begin{figure}
	\includegraphics[width=\linewidth]{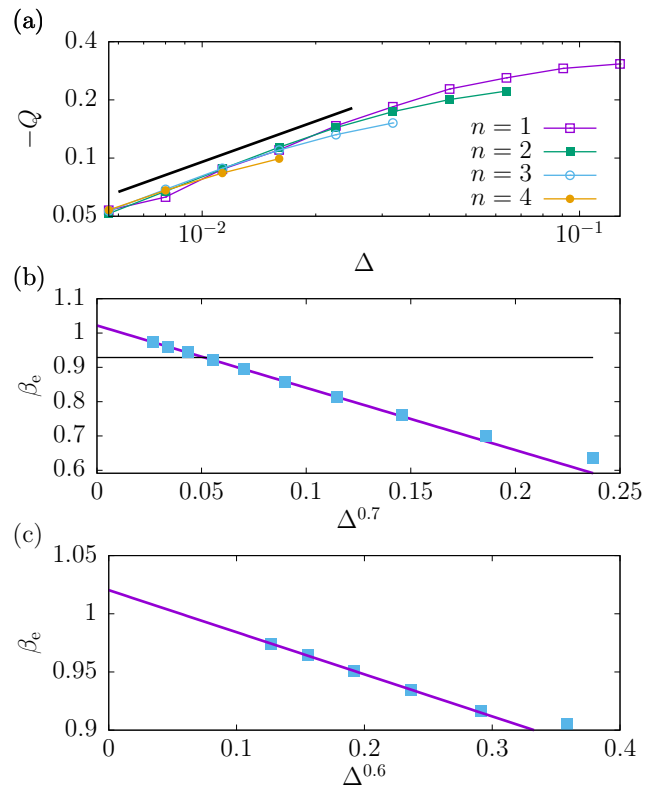}
	\caption{\label{Fig:B4} (a) Plots of $-Q$ vs $\Delta$ for $b=2^{n/2}$ with
	$n=1$, 2, 3, 4 (top to bottom) on a double logarithmic scale
	for the BAW(4).
	The line segment with slope 0.7 is to guide to the eyes.
	(b) Plot of $\beta_\text{e}$ vs $\Delta^{0.7}$ with $b=2$ for the BAW(4). 
	The horizontal line indicates the value $\frac{13}{14}\approx 0.93$. 
	The extrapolation gives $\beta = \betae$.
	(c) Plot of $\beta_\text{e}$ vs $\Delta^{0.6}$ with $b=2$ for the BAW(2). 
	The extrapolation gives a result consistent with the BAW(4).
	}
\end{figure}

To find $\beta$, 
we first analyze $Q$, defined in Eq.~\eqref{Eq:Q}.
First of all, we would like to mention that $Q$ is found negative, 
which indicates that $B$ in Eq.~\eqref{Eq:scale_rhos} is negative.
Actually, we found $B\approx -2$.
Accordingly, $\beta_\text{e}$ should approach the ordinate from below;
see Eq.~\eqref{Eq:betae}.
In Fig.~\ref{Fig:B4}(a), we depict $-Q$ as a function of $\Delta$
for $b = \sqrt{2}, 2, 2\sqrt{2}$, and $4$. 
In this plot, $\chi$ is set to be 
0.7, which results in the asymptotic behavior of $Q$ being independent 
of $b$. Also, this choice of $\chi$
is consistent with the asymptotic behavior
of $Q$ for small $\Delta$; see the line segment with slope $0.7$ in Fig.~\ref{Fig:B4}(a).

Since we have found $\chi$,
we can now analyze the effective exponent $\beta_\text{e}$ efficiently.
In Fig.~\ref{Fig:B4}(b), we plot $\beta_\text{e}$ with $b=2$ against
$\Delta^{0.7}$, which shows a nice straight line behavior
for small $\Delta$. As anticipated, $\beta_\text{e}$ indeed decreases with $\Delta$. By a linear extrapolation, we find
$\beta\simeq \betae$, where
the number in parentheses indicates uncertainty of the last digit. 
For comparison, we also draw the conjectured value $\frac{13}{14}$
in Fig.~\ref{Fig:B4}(b) as a horizontal line. 
Obviously, the effective exponent goes beyond 0.93, which 
shows that the conjecture about $\beta$ cannot be correct.
We would like to mention that other choices of $b$ give a 
consistent conclusion (details not shown here).

We also studied the behavior of the steady state density
for the BAW(2) with $q=0.5$. The critical point of this 
case is available from Ref.~\cite{Park2013}, which is
$p_c \simeq 0.494~675(25)$. We use this value for the following
analysis.
We simulated the BAW(2) at $ p = p_c -\Delta_n$
for $n=10,11,\ldots,17$ with system size $L=2^{20}$.
The number of independent runs ranges from 1000 to 2000.

We first analyzed $Q$ to obtain $\chi = 0.6$ (details not shown here).
Since the exponent of the leading behavior of corrections to scaling 
is not universal, it is not surprising that $\chi$ of the BAW(2)
is different from that of the BAW(4).
With this estimate of $\chi$, we also analyzed $\beta_\text{e}$,
which is shown in Fig.~\ref{Fig:B4}(c). A linear extrapolation
gives an estimate $\beta \approx 1.02$, which is consistent with
the estimate for the BAW(4). 

\begin{figure}
	\includegraphics[width=\linewidth]{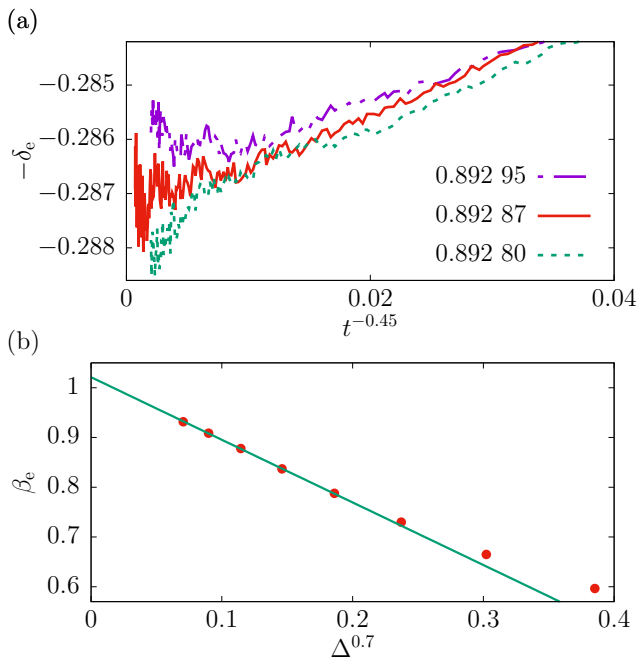}
	\caption{\label{Fig:im} (a) Plots of $-\delta_\text{e}$ vs $t^{-0.45}$ for the CP2 with $b=16$.
	The values of $\lambda$ are $\lambda=0.8928$, 0.892~87, and 0.892~95 (bottom to top).
	$-\delta_\text{e}$ for $\lambda=0.892~87$ exhibits a linear behavior,
	while $-\delta_\text{e}$ for $\lambda=0.892~95$ (0.8928) veers up (down) for large $t$,
	which gives $\lambda_c \approx 0.892~87$.
	(b) Plot of $\beta_\text{e}$ vs $\Delta^{0.7}$ for the CP2 with $b=2$. 
	The result of a linear extrapolation for small $\Delta$ is drawn as a straight line, which gives $\beta\approx 1.02$.
	}
\end{figure}
Now we move on to the CP2.
As we have already noted, the CP2 is identical to the interacting monomers model~\cite{PP2008PRE}. The critical point
$\lambda_c$ is available from Ref.~\cite{PP2008PRE}, but this value is less accurate because
the analysis was based on the conjectured value of $\delta$. So we will find 
the more accurate critical point by analyzing $-\delta_\text{e}$.

The simulated system size is $2^{23}$ and the longest observation time (for $\lambda=0.892~87$) is $T_{309}$. $\rho(t)$ is obtained by
averaging over 1000 independent realizations.
We first found the exponent of corrections to scaling to be $0.45$ (details not shown here).
Using this value, we plot the effective exponent 
$-\delta_\text{e}$ as a 
function of $t^{-0.45}$ in Fig.~\ref{Fig:im}(a). 
From this analysis, we conclude that 
the critical point is $\lambda_c = 0.892~87(8)$,
which is indeed different from that found in 
Ref.~\cite{PP2008PRE}. The critical exponent $\delta$ is found to be around 0.2872.

To obtain $\rho_s$, we simulated a system of size $L=2^{20}$ at $\Delta \equiv \lambda -\lambda_c =\Delta_n$ with $n = 9, 10, \ldots, 18$.
For each case, 1000 independent runs are averaged. 
We first estimated $\chi$ to be 0.7 (details not shown here),
which is used to analyze $\beta_\text{e}$.
In Fig.~\ref{Fig:im}(b), we depict $\beta_\text{e}$ against 
$\Delta^{0.7}$. Again by a linear extrapolation, we 
obtain $\beta \approx 1.02$. 

Since the numerical estimation of critical exponents is influenced by the
accuracy of the critical point, we have to check how much the effective exponent $\beta_\text{e}$
is affected by the error of the critical point.
Let $\pi_c$ be the ``exact'' critical point and $p_c$ be the numerically found critical point. 
When $p$ is close to $\pi_c$ in the active phase, the steady state density is well approximated by
\begin{align}
\rho_s(p) \simeq A |\pi_c-p|^\beta \left ( 1 + B |\pi_c-p|^\chi  \right ),
\end{align}
where we have kept terms only up to the leading correction to scaling.
If we expand $\rho_s(p)$ around $p_c$ (rather than the exact $\pi_c$), we get
\begin{align}
	\nonumber
	\rho_s(p) &\approx A |\Delta + D_c|^\beta \left [ 1 + B |\Delta+D_c|^\chi \right ]\\
	&\approx A \Delta^\beta \left [ 1 + B \Delta^\chi + \beta \frac{D_c}{\Delta}
\right ],
\end{align}
where $D_c \equiv |\pi_c-p_c|$, $\Delta \equiv |p_c-p|$, and we have kept terms up to $D_c/\Delta$.
Notice that $D_c$ is smaller than $\Delta$ in our simulations.
As in Sec.~\ref{Sec:result}, we find the effective exponent 
\begin{align}
	\beta_\text{e}(\Delta;b) \approx \beta + B \Delta^\chi \frac{b^\chi-1}{\ln b}
	- \frac{b-1}{b \ln b} \frac{D_c}{\Delta}.
\end{align}
For $b=2$, the largest error (for smallest $\Delta_\text{min}$ in simulations) 
due to the inaccuracy of the critical point is
\begin{align}
	\frac{2-1}{2\ln 2} \frac{|D_c|}{\Delta_\text{min}}
	=
	\begin{cases}
		5\times 10^{-3},& \text{BAW(4)},\\
		6\times 10^{-4},& \text{BAW(2)},\\
		3\times 10^{-3},& \text{CP(2)},
	\end{cases}
\end{align}
which is of size similar to the statistical error.
Hence, the inaccuracy of the critical point in our simulations hardly changes the estimate.

We have studied three different cases and we consistently obtain $\beta=\betae$.
The estimated $\beta$ in this paper along with results
in Ref.~\cite{Park2013} also gives $\nu_\perp = \nupee$ and
$\nu_\| = \nupae$.

\section{\label{Sec:sum}Summary and discussion}
To summarize, we extensively studied a few models that
belong to the directed Ising (DI) universality class,
focusing on estimating the order parameter exponent $\beta$.
We first analyzed the corrections-to-scaling function $Q$ defined in Eq.~\eqref{Eq:Q}. 
After finding the asymptotic behavior of the corrections-to-scaling function 
for each model, we analyzed the effective exponent $\beta_\text{e}$, to arrive at
$\beta = \betaer$, which is clearly different from the widely
acknowledged value $\approx 0.93$.
Using the estimated value $\beta$ along with the estimates of other
exponents in Ref.~\cite{Park2013}, we also arrived at 
$\nu_\perp =\nupee$ and $\nu_\|=\nupae$, which are larger than the 
conjectured values in Ref.~\cite{J1994}. Critical exponents are 
summarized in Table~\ref{Table:di} with comparison to the conjecture.

\begin{table}
	\caption{\label{Table:di}Critical exponents of the DI class in one dimension.
	Numbers in parentheses indicate uncertainty of the 
	last digits.}
	\begin{ruledtabular}
		\begin{tabular}{lll}
			Exponent&Conjecture~\cite{J1994}&Numerical values\\
		\hline
		$\beta/\nu_\|$&$\frac{2}{7}\approx 0.2857$&0.2872(2)\\
		$\beta/\nu_\perp$&$\frac{1}{2}$&0.5000(6)\\
		$z$ & $\frac{7}{4}=1.75$&1.7415(5)\\
		$\eta$ &0&0.0000(2)\\
		$\beta$ &$\frac{13}{14}$&\betaer\\
		$\nu_\perp$ &$\frac{13}{7}\approx 1.86$&\nupee\\
		$\nu_\|$ &$\frac{13}{4}\approx 3.25$&\nupae
		\end{tabular}
	\end{ruledtabular}
\end{table}
Previous Monte Carlo simulations have repeatedly found that $\beta$ is around 0.93 (though its value is somewhat scattered).
As we have shown in this paper, however,
corrections to scaling are not negligible for the BAW($2m$) and
the CP2. Thus, without taking them into account, one can be
easily misled by Monte Carlo simulations.
Interestingly, our result is consistent with the analysis of
series expansions~\cite{Jensen1997, InuiT1998PRL}.
It is also interesting that nonperturbative renormalization group analysis 
gives $\nu_\perp = 2 \pm 0.1$~\cite{Canet2005}, which is comparable to our estimate $\nu_\perp = \nupee$.

\begin{figure}
	\includegraphics[width=\linewidth]{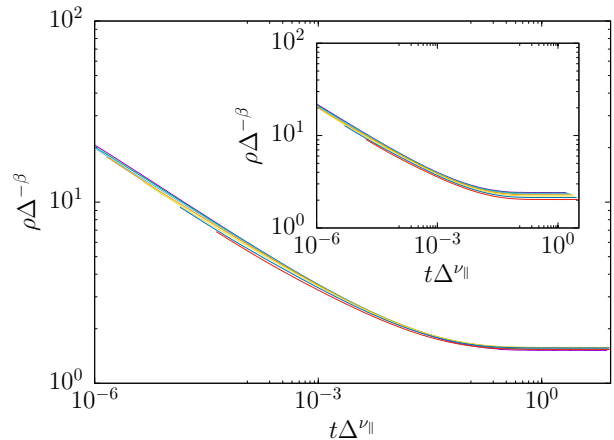}
	\caption{\label{Fig:SC4} Scaling collapse plot of $\rho \Delta^{-\beta}$ vs $t\Delta^{\nu_\|}$ for the BAW(4). Here $\Delta=\Delta_n$ with 
	$n=5,6,\ldots,11$ and we use $\beta = 0.93$ and $\nu_\|=3.25$.
	Inset: Same scaling collapse plot with $\beta = 1.02$ and $\nu_\|=3.55$.
	Even though the exponents are more accurate, the scaling collapse looks
	worse, which should be attributed to the corrections to scaling
	not to the value of the exponents.
	}
\end{figure}

Many studies used either a power-law 
fitting~\cite{J1994,Zhong1995,Argolo2020} 
or a scaling collapse~\cite{KP1994}.
To illustrate what we could have concluded if we had not taken corrections
to scaling into account, we present a scaling collapse for the BAW(4) in
Fig.~\ref{Fig:SC4} using two sets of exponents.
As one can see, the conjectured exponents give a better-looking scaling collapse than
the more accurate values (see the inset of Fig.~\ref{Fig:SC4}). 
Since the conjectured values give an ``impressive'' scaling collapse,
a naive power-law fitting to our data would have supported the conjecture.

The reason why a scaling collapse does not look perfect with the accurate
values of $\beta$ and $\nu_\|$ can be explained by 
the effective exponent in Fig.~\ref{Fig:B4}(b). The largest value of $\beta_\text{e}$
we have is still somewhat away from the extrapolated value. 
On this account, setting $\rho_s \simeq A \Delta^{1.02}$ for our data 
is an inaccurate approximation. Using $B \approx -2$ and $\chi \approx0.7$  for the BAW(4)
obtained in Sec.~\ref{Sec:result}, 
we find that $\beta_\text{e}$ becomes larger than 1 for $\Delta < 0.0016$.
For the system with $\Delta=0.0016$, the steady state will be attained after 
$t \approx  10^{10}$.
Thus, it would be very difficult to have a nice-looking scaling 
collapse with the correct exponents for models belonging to the DI class.

This discussion adds a caveat to numerical analyses of critical phenomena. 
A nice-looking scaling collapse does not 
mean that the critical exponents are accurately found and, by the same token,
a less impressive scaling collapse should not be a reason 
to reject the possibility of there being critical exponents. 
A scaling collapse can at best be used to check consistency;
whether a scaling collapse is nice looking or not
is too subjective. A scaling collapse is reliable
only when corrections to scaling are negligibly small, which one cannot be sure of
\textit{a priori}. In the same context, a naive power-law fitting is a misleading practice when it
comes to determining the critical exponents. 

The effective exponent was actually studied in Refs.~\cite{HO1999,MO2000},
which predicted $\beta = 0.95(2)$. However, the numerical error of the effective
exponent in Refs.~\cite{HO1999,MO2000} is quite large.
Besides, $\chi$ was set to 1 without any numerical support in the analyses of the effective exponent 
in Refs.~\cite{HO1999,MO2000}.
On these accounts, we claim that our estimate is more accurate than that in the former
studies~\cite{HO1999,MO2000}.

It might look strange that our estimate of $\beta$ is larger than the order parameter exponent of the 
mean-field theory, which is 1. However, this observation is not completely inconsistent with
the field theory~\cite{CT1996,CT1998}. The renormalization-group calculation up to one loop order
suggests that $\beta \rightarrow \infty$ as dimension $d$ approaches $d_c'=\frac{4}{3}$ from above~\cite[see page 20]{CT1998}. In this context, it is not impossible that $\beta$ in lower dimensions is larger than the corresponding
value in the mean field theory.
Nonetheless, it is hard to rule out that $\beta$ is actually 1 and, accordingly, $\nu_\perp=2$.
If this is indeed the case, the DI class seems to have (at least) three integer critical 
exponents, $\eta=0$, $\beta=1$, and $\nu_\perp=2$. Since $\beta$ and $\nu_\perp$ describe steady-state behavior,
solvability of the DI class in one dimension, probably only for the steady state, might be still open (notice that $\beta$ and $\nu_\perp$ of the two-dimensional Ising model are fractional numbers, whereas dynamics exponent $z$ is presumably not; for a recent accurate 
estimate of $z$ of the two-dimensional Ising model with Metropolis algorithm, see Ref.~\cite{My2015JSTAT}). 
It would be an intriguing future project to come up with a theory that explains whether $\beta$ is exactly 1 or not.

\begin{acknowledgments}
	This work was supported by the Basic Science Research Program through the
National Research Foundation of Korea~(NRF) funded by the Ministry of
Science and ICT~(Grant No. 2017R1D1A1B03034878), and by the Catholic University of Korea,
research fund 2018. 
The author furthermore thanks the Regional Computing Center of the
University of Cologne (RRZK) for providing computing time on the DFG-funded High
Performance Computing (HPC) system CHEOPS.
\end{acknowledgments}
\bibliography{Park}
\end{document}